# Stream Processing for Solar Physics

Applications and Implications for Big Solar Data

Karl Battams
Space Science Division
U.S. Naval Research Laboratory
Washington, D.C.
karl.battams@nrl.navy.mil

*Abstract*— Modern advances in space technology have enabled the capture and recording of unprecedented volumes of data. In the field of solar physics this is most readily apparent with the advent of the Solar Dynamics Observatory (SDO), which returns in excess of 1 terabyte of data daily. While we now have sufficient capability to capture, transmit and store this information, the solar physics community now faces the new challenge of analysis and mining of high-volume and potentially boundless data sets such as this – a task known to the computer science community as stream mining. In this paper, we survey existing and established stream mining methods in the context of solar physics, with a goal of providing an introductory overview of stream mining algorithms employed by the computer science fields. We consider key concepts surrounding stream mining that are applicable to solar physics, outlining existing algorithms developed to address this problem in other fields of study, and discuss their applicability to massive solar data sets. We also discuss the considerations and trade-offs that may need to be made when applying stream mining methods to solar data. We find that while no one single solution is readily available, many of the methods now employed in other data streaming applications could successfully be modified to apply to solar data and prove invaluable for successful analysis and mining of this new source.

## I. INTRODUCTION

Beginning at the advent of space-based scientific investigation in the 1950's, increasingly complex and advanced instrumentation has routinely been launched in to space to study the Sun. As imaging, data storage and data transfer technology has advanced, we have witnessed an ever-growing volume of solar physics data with each new mission launched. The P78-1/SOLWIND solar coronagraph, for example, returned a total of around 1GB of imaging data during its period of operation from 1979–1982. The Solar and Heliospheric Observatory (SOHO), operational since 1995, returns this same volume of data approximately every two days, with a total mission data volume of a few terabytes. The Solar Terrestrial Relations Observatory (STEREO), operational since 2007, was at peak returning a few gigabytes daily, with a total mission volume in the tens of terabytes.

Until recent years, this steady rise in data volume has largely remained commensurate with available storage capabilities and desktop computer processing power. However, the launch of the Solar Dynamics Observatory (SDO) in 2010 has set a new pace for solar physicists. Returning approximately 1.5 terabytes of solar imaging data daily, the SDO satellite presents a watershed moment for solar physicists and the way they must handle, process, analyze and mine an unprecedented volume of data. To realize the full potential of this data, the community must look beyond desktop processing methods employed for previous data sets and instead seek solutions from computational disciplines that are more familiar with such data volume and velocity.

In this survey paper, we approach this problem from the computational field of *data stream mining*, considering how techniques and methods used in this field may apply to the special case of analysis and mining of big solar data. Several of the methods discussed here are drawn from the Data Stream: Models and Algorithms survey [1], with the most pertinent and promising of the methods and algorithms singled out and evaluated in the context of the big solar data problem. Potential solar physics applications of the methods are discussed, along with modifications, adaptations and limitations that the community should consider.

We begin with an overview of data streams, defining the terms and outlining the challenges that they pose. We then discuss briefly the nature of solar data and, in particular, the SDO data that drives this paper. Then we discuss several specific aspects of stream mining that we feel most relevant to the big solar data problem: *Classification and Clustering*; *Time Horizons*; *Synopsis Methods*; and *Change Detection and Concept Drift*. We consider each in the context of solar data, discussing the advantages and limitations of each. In many cases, examples of existing algorithms are offered as a starting point for future exploratory works in this emerging field.

## II. DATA STREAMS

A data stream is considered to be an unbounded or infinite incoming stream of data, and stream mining the computationally intensive process in which stream data is mined and/or analyzed in real-time. Stream data can arrive in many forms, and from many sources. Examples include remote sensor data (e.g. temperature, pressure), Internet network traffic, financial transactions or stock exchanges, video or still-frame images (e.g. CCTV, drones), satellites, cell phones, cars, and numerous other sources. Some of these streams simply by themselves generate huge volumes of data, but in aggregate, the data volume becomes truly boundless. We are now at a stage where our ability to *capture* data can begin to outweigh our ability to *process and analyze* that data.

Data streams pose numerous problems for data analysis and mining. Given their boundless nature, it is not feasible to mine from the entire archive. Instead, methods must operate on either just the most recent data element, or a subset of the data within some moving time window, $t$. In the latter approach, careful consideration must be given to the size of the window to keep within the computational limits of the system performing the mining. Algorithms are presented later in this paper that weight the value of old versus new data, for example, or sample older data at a lower rate.

The nature of stream data is typically such that many mining algorithms can only afford *one pass* at the data before a new element arrives. In streams with fluctuating stream velocity, it may also sometimes be necessary to perform *load shedding* [6] to omit some of the incoming data if the volume becomes overwhelming. For the solar physics case, the incoming data rate is usually uniform and thus we do not consider load shedding further in this paper. However, future missions may indeed pose this problem as they go through alternate period of high and low data rates.

Popular methods of stream mining include distributing the computation over multiple nodes, or reducing or summarizing the data into a more reduced form (*data synopsis*). Methods that traditionally operate on entire static datasets, such as clustering or frequent pattern mining often require significant modifications if they are to be used in this context [12]. In part, this is due to the potential for *concept drift* or sudden changes in the nature of the stream, whereby previously held assumptions about the data, or pre-determined models of the behavior, no longer fully apply.

While the sources and applications of stream data and stream data mining are numerous, we focus our attention on a single case of a new data stream that has appeared at the forefront of science research: the NASA Solar Dynamics Observatory, or SDO. This mission returns an unprecedented 1.5Tb of new data *daily*, representing an unprecedented challenge to scientists familiar only with the small to moderate volumes of imaging data.

### III. THE CHALLENGE OF NASA'S SOLAR DYNAMICS OBSERVATORY

The past four years have been transformative for the solar physics community following the launch of the NASA Solar Dynamics Observatory[1] in February 2010. This state-of-the-art space observatory boasts three ultra-high resolution cameras that observe the Sun's outer atmosphere (corona), its surface and its magnetic field in unprecedented detail. Images are returned from this satellite at 4096x4096 pixel resolution, and at an average rate of one every 10-12 seconds on an ongoing basis from each instrument, with an effective data rate of a 32MB image every few seconds. The result is approximately 1.5Tb of new images daily in several wavelength channels over the three instruments.

The introduction of this capability is an enormous boost to space-based studies of the Sun, which until now had sufficed with around 500MB of daily data from SOHO[2] and a few gigabytes per day from the twin STEREO[3] satellites, distributed over numerous imaging instruments. Prior to SDO, this combined daily data load from SOHO and STEREO provided a wealth of data about the Sun and its transient events such as coronal mass ejections and solar flares, but in a manageable volume. The combined data stream from these missions is sufficiently small that assembling a day or two of images into a data cube or animated sequence for interpretation and analysis is a straight-forward task with minimal computational requirements, and analyzing several days at a time is not overly challenging.

The SDO data stream, however, provides an entirely new challenge to a science community that has previously relied largely on its ability to access and process large temporal spans of data at any one time. With SDO's extremely high spatial and temporal resolution, it is no longer feasible to use our existing methods to analyze every image returned. One approach to this problem is to treat observations as elements in data stream, and then look to established methods used within the computer science community for handling such data. While community emphasis is focused upon the SDO mission data, both current and future ground and space-based observatories will undoubtedly be returning volumes of data that can be treated as proposed in this paper.

### IV. EXISTING SOLAR DATA MINING WORK

The application of data mining methods to solar data is not new, with several existing efforts aimed at automatically detecting specific features in solar data. Solar feature detection and classification is central to the Heliophysics Event Knowledgebase (*HEK*) – a freely available online repository of solar events accessible via the Virtual Solar Observatory[4] (*VSO*) – and is part of the broader SDO Event Detection System (EDS) effort [18] The EDS looks to analyze solar data in near real-time, incorporating "feature-finding modules" contributed by the community. A comprehensive list of current algorithms is given by [18] and include methods designed to detect coronal holes, sunspots, flares, and EUV waves, to name a few. Some of these algorithms run continuously on incoming data whereas others are event-driven, triggered by higher-level detection routines. The HEK team actively solicits new codes for submission to the program.

Beyond the SDO data, the Computer Aided CME Tracking (CACTus) algorithm, [24] is a tool designed to automatically detect coronal mass ejections (CMEs) in solar coronagraph data. Similar efforts such as ARTEMIS [9] and SEEDS [21, 22] have also been developed, though all returned slightly different classification results when compared side-by-side [9].

---

[1] http://sdo.gsfc.nasa.gov
[2] http://sohowww.nascom.nasa.gov
[3] http://stereo.gsfc.nasa.gov
[4] http://www.virtualsolar.org

Clearly the community is aware of the challenge of mining solar data, and the need for advanced algorithms to address this problem. Programs are in place to do this, but it remains an emerging field of study for the solar community, still somewhat in its infancy. The following section of this paper looks broadly at the problem of stream mining from the computer science perspective, and how methods employed by that discipline might apply to the solar physics case.

## V. STREAM MINING AS APPLIED TO SOLAR DATA ANALYSIS

Before entering discussion of stream processing methods that might apply to SDO data, we first outline the nature of the SDO stream, and what we might hope to achieve through mining the data, noting that conclusions drawn here are generally applicable to all sources of solar data. Stream mining is not a one-stop solution for all analysis efforts in this field, represents a practical and viable solution when applied appropriately to certain subset of problems.

SDO returns images in real-time or near real-time, at a cadence of around 2-10 seconds, depending on the channel (filter wavelength) under scrutiny. While sometimes this rate my drop due to special observations or operational constraints, for all practical purposes there exists a maximum data rate of one 4096x4096 pixel image every two seconds – or approximately 16MB per second. This provides a useful upper bound to which mining algorithms could be catered without further consideration of fluctuations in the stream.

The data returned are images of the super-heated solar atmosphere or the solar surface, and while the wavelength ranges may vary, the fundamental data type itself remains consistent and from just one source (as opposed to systems that may try to aggregate, say, imaging and sensor data from numerous sources). Again, we have an advantage with the predictability and uniformity of the data type.

At any given time, the solar surface contains a number of categorized features that may be of interest, and exist on differing timescales. So-called "*spicules*" are small jets typically seen at the solar poles that may last few tens of minutes to hours. *Coronal holes* are large-scale dimming features within the extreme ultraviolet (EUV) solar corona that at times can occupy up to 50% of the visible face of the Sun. These tend to last for days, but have boundaries that evolve on timescales of several hours. *Active regions,* associated with sunspots on the visible solar surface, are large areas of heightened magnetic field strength, complexity and energy, and are the source regions for solar flares and many coronal mass ejections (CMEs). They can persist for more than a 27-day solar rotation, but the eruptive events themselves – CME's and flares – are short duration events with the latter in particular occurring on the timescale of minutes.

We now have unprecedented temporal and spatial resolution, and answers to many of these questions likely exist within this new data set. However, it may also be that crucial clues to these problems are somewhat hidden within the data set, with traditional analysis methods under-sampling the data. There may also be new, short-duration periodic features that remain undetected simply because they have not been looked for, or statistical patterns or periodicities that may reveal themselves with appropriate analyses.

Methods employed by the solar community are often event-driven and rarely is every pixel in every image considered within even a short time window. To maximize the return from big solar data, the community ideally needs to consider methods that offer the capability to sample every pixel in every image if needed, assessing them in terms of solar features and the statistical properties of the data. To achieve this, a number of techniques will need to be employed, including some combinations of clustering, classification, pattern mining and data synopsis methods. These techniques alone will not address all questions, but may offer the opportunity to study previously unconsidered aspects of the data.

Now follows a discussion of the data streaming tools, methods, algorithms and general considerations that could be applicable to analysis of *solar stream data*. This survey covers algorithms applicable mainly to basic data mining methods, the utility of which will depend on the nature of mining task.

### A. Classification and Clustering

Fundamental to many data mining applications is *classifying* or *clustering* the data into classes that meet some defined criteria of "similarity". Typically this means either categorizing the data at it arrives, or categorizing certain features within the data following some feature extraction step. In the case of analyzing the SDO solar data, for example, we may want to automatically flag images as containing a coronal hole, a certain magnetic field configuration within an active region, or perhaps something more complex such as a known precursor to flare.

A fundamental motivation for SDO is to provide visual imagery of the solar surface so that we can understand and monitor its behavior and in particular track features that may ultimately affect Earth. These would include active regions that spawn solar flares, or coronal holes that are the source of very fast solar wind streams. These are typically large-scale and reasonably well-defined features, but it may also be the case that small, transient features are located and observed, and it would be of importance to catalog and categorize these features. In a small data volume, simple visual inspection or crude feature extraction algorithms would be sufficient, but for SDO's high-volume stream requires a different approach.

The method of *clustering* enables the collecting and cataloging of similar data elements or features. These may be known features that the algorithm is "trained" to detect, or unknown features that the algorithm decides are similar in some way. Traditional clustering methods do not work well with stream data, as they prefer to see the entire data set together and categorize based on the entire volume. With stream data, only one or a few elements can be considered at any given time, and often the algorithm is unable to compare current clusters to older archived clusters.

One suggested approach is the CluStream [2] algorithm which separates the clustering process in to two parts: an

"*online micro-clustering*" and "*offline macro-clustering*" approach, where the offline clustering takes advantage of the summary statistics recorded from the online clustering. CluStream performs so-called *micro-clustering* – a process that maintains the statistics of the stream at a sufficiently high-resolution such that it is later useful to the offline macro-clustering algorithm. The macro-clustering will then look to classify based on the summarized micro-clusters.

Micro-clusters are statistical summaries that contain similar information to *cluster feature vectors* [1] but with additional metadata such as timestamps. An advantage of micro-clusters is that they are able to capture the evolution of the stream as well as provide a good summary of the incoming data. (Data evolution is discussed further in Change Detection and Concept Drift.) A critical element to capturing the micro-clusters is to record them at appropriate intervals, for which a *pyramidal* scheme [2] captures micro-cluster snapshots at different levels of granularity depending on the age of the data. This is discussed further in the Time Horizons section.

Establishing and maintaining micro-clusters requires a fast-paced algorithm that assesses each data point as it arrives and determines whether it should belong to an existing cluster or become a new cluster of its own. Clustering decisions are based upon the distance of the data point from existing clusters. (Prior familiarity with distance metrics used in data mining and cluster analysis are assumed in this paper, with myriad resources online[5] and in print [16] available as a primer in this subject.) When a new point is found to be within a pre-determined *maximal boundary factor*, it is merged with existing points in that cluster. Otherwise the point becomes the basis for a new cluster. In the latter case, however, the new formation of a cluster means that the overall number of clusters must be reduced by one – something that is achieved by either retiring an "old" cluster (based on the average age of the cluster), or by merging two clusters together.

A search of literature reveals numerous alternative clustering methods that have been successfully applied in the stream mining context, such as STREAM for *k*-Median clustering [14], HPStream [2] for higher dimension data streams, and E-Stream [28] for efficient handling of change and evolution in discovered clusters.

Related to clustering data is classification, where the goal is to apply certain labels or classes to features once unique groupings or clusters have been established. For solar data, classification would be an important step for identifying new features, and labeling them for later queries. A detailed discussion of a method for classifying stream data is made by [2] who propose an "*on demand classification process*" that uses micro-clusters to refine the best time-horizon for classifying the data. This method uses a subset of the micro-clusters from the classification training stream to fine-tune the time-horizon that will be used on the actual stream itself. To address concept drift in data streams, [19] propose an ensemble method that enables automatic detection of new classes as they arrive while circumventing the need for an extensive time delay on classifying the new data class. Alternative approaches are listed in [27], [30], and [12].

One advantage with the NASA SDO data stream is that the fundamental nature of the data is unchanging – i.e. the data are images of the Sun – and thus a subset of the archived data can be readily obtained at any time for training and tuning of algorithms to be used on the incoming stream itself. However, this learning in itself is a challenge when faced with a data stream, and modifications to conventional approaches are necessary. One suggested method [11] utilizes so-called *Hoeffding Trees* – a decision-tree based method that does a good job of balancing the trade-off between accuracy and efficiency of models produced. Decision-tree methods [26] are popular for classification problems due to their efficiency and accurate and easily interpreted results, while being capable of handling simple or complex multi-dimensional data [16]. Hoeffding trees expand on this idea by providing trees that can learn in real-time, with improving accuracy as more data flows in through the stream, and a provable guarantee that the results obtained online and in real-time differ minimally from those obtained offline in the so-called batch mode.

The Hoeffding Tree method can be further extended to a system known as *Very Fast Decision Trees*, or VFDT [11] that looks to overcome some of the Hoeffding limitations by better addressing issues surrounding ties of attributes and bounded memory constraints. For example, the VFDT algorithm allows for delaying of classification decisions in ambiguous cases until the ambiguity can be better relieved by following a user-defined error bound.

The "*Adaptive Nearest Neighbor Classification for Data-streams (ANNCAD)*" [13] takes advantage of *Haar wavelet transforms* [4] (discussed in the Synopsis Methods) by representing the data at different levels of granularity or resolution, while attempting to perform classification. If a suitable classification is not found with some resolution then the grid size is changed and the classification attempted again. Such methods offer tremendous advantage as the data size is reduced significantly to more manageable levels, and then increased only as needed to meet requirements of the processing method. This general technique is one that can be applied to many aspects of data and stream mining, and is not necessarily limited to classification or cluster analysis.

Related to clustering and classification is the problem of frequent pattern mining, where the goal is to specifically locate recurring features within a data stream. For static data sets, this is not a difficult task as the data can be considered in whole, and all patterns analyzed to see which are the most frequent as a result of several passes over the data set. With stream data, the "memory" of previous frequent patterns can still be retained but the algorithm needs to both adapt to new patterns that may arise as the stream changes and evolves and – crucially – adapt to the one-pass nature of stream data. This requires a robust and computationally efficient algorithm that is capable of making the compromise between efficiency and the accuracy of results. Several such algorithms are detailed by [3] though in many of these cases, it is somewhat implied that the items or features to be counted are known *a priori*.

---

[5] https://en.wikipedia.org/wiki/Cluster_analysis

This could become a problem if new and undetected features arise that have not been accounted for. Much of this deficiency may be alleviated by use of appropriate adaptive clustering methods and change detection algorithms that are able to account for *concept drift* in the data.

If feature detection, extraction and tracking is a desired goal of stream processing solar data, then clustering and classification will be a fundamental component of that analysis. Solar features such as active regions will tend to evolve over periods of days or more, with small-scale locale dynamics such as flares occurring on much shorter timescales. Thus, algorithms that can capture short-duration, sporadic events are needed, while maintaining the long-term tracking capability over a longer time-horizon. The micro-clustering approach described here, combined with an appropriate learning algorithm, would seem to be a good starting approach.

### B. Time Horizons

Two general approaches exist to address the boundless nature of data streams, which means it is impossible to address the entire data set at one time. The first is to simply only consider the current data element and drop it from memory as soon as the next arrives. The second approach is to retain as much recent data as possible within the system constraints in some predetermined "time window". The choice of this *time horizon* is constrained by several factors including system memory limitations, computational resources, and the nature of the mining algorithm.

For certain mining tasks, the first approach, where only the current data element is considered, may be sufficient. Simple counting problems, for example, only need store a counter variable and not a subset of the data, and thus only the current data element is of interest. But for many problems, including those of classification or clustering, a subset of recent data needs to be retained in memory in order to obtain accurate summaries, synopses, or classification models.

Probably the most basic time-horizon model is a simple sliding window, where a subset of the most recent data is retained either based on a fixed number of data elements, or a fixed time period, with the system memory and computational capabilities being the primary constraint. As a new data element arrives, the oldest element is simply discarded, and the window progresses in time with the data. In the solar data consideration, a full-resolution image is 32MB and for a typical to moderate system with, say, 32GB available RAM, we can crudely estimate that approximately 1,000 images could be stored at one time. With SDO, for example, returning a new image every two to ten seconds, this is only corresponds to around 1 – 3hrs of the most data in memory at any given time. This may be sufficient for short-duration events or structures, but is clearly insufficient for long-term trends. This simplistic view also does not account for the ability of the system to actually mine a 32GB data cube, but illustrates the limitations of a time window even under ideal circumstances.

Thus in the solar data circumstance, and certainly many others, an alternate scheme is needed in which some degree of historical data is retained, but at a lower resolution than the newest data. The *pyramidal time window* [2] algorithm is an excellent example of a method that retains data over a long window in time, but with the older data retained at a much lower level of granularity. With appropriately chosen time-intervals, it is demonstrated [2] that data values between certain recorded moments, or *snapshots*, can be well approximated by using the two closest snapshots to the desired data element, and that no single snapshot is more than a factor of two away from the specified time horizon.

Similarly, the *tilted timeframe model* [15] relies on storing older data at a lower level of granularity. Within this class are three subset models, known as the *natural, logarithmic,* and *progressive logarithmic* tilted time frame models. The natural model creates the granularity on a natural timescale (e.g. weeks, days, hours, minutes); the logarithmic model stores data at times *t, 2t, 4t, 8t, 16t,* etc; and the progressive logarithmic sees snapshots placed into different frame numbers with a user-defined maximum number of snapshots being held at any given level.

In some instances it may be that only the most recent hour or two of data is required and in fact only the very most recent $n$ data elements are of primary interest, with older data weighted linearly or exponential with less importance (known as a *damped window* [17]). Here, scaling factors can be used to decrease the weight of older snapshots or data elements, either based upon a linear or exponential scale.

Finally, some situations may require the time window to be adaptive to the task being solved, and optimized in length during the computation. One such scheme is the previously mentioned micro-clustering method, that uses a subset of training data from the stream to tune its parameters and pick the optimum time horizon that minimize the error in results while keeping within computational limits.

For solar data analysis, the time horizon chosen, and the model used for maintaining any snapshots, will be highly dependent on the goals of the analysis and mining task. In particular, attention must be paid to whether all recent data are of equal importance or value, or if the most recent elements are more important than older elements. For example, a simple method for automatic solar flare detection would only need to consider the most recent several minutes of data in order to capture and flag an event. However, for long-term evolutions it may be that data is required encompassing a half solar rotation as a feature transits the disk. Combining data streams from other instruments (e.g. STEREO) could conceivably require yet longer windows if indeterminate length, with the upper bound be marked by the disappearance of the tracked feature. Given the large size of the SDO data files, maintaining a significant volume in memory becomes unwieldy and computationally impractical to maintain, and some method for reducing this data volume while maintaining the quality of the data. For this, data summary and synopsis methods will play an important role.

### C. Synopsis Methods

In many instances, a data volume in a stream may be too large for practical analysis on the element itself, and some means of reducing the size, dimensionality or complexity of the data is necessary. For solar data, it is frequently the case that only a certain percentage of the pixels or certain signals within the data are of value, and the rest a drain on computational resources for any given mining task.

This is where data reduction and synopsis methods come into play, as they look to reduce the amount of data to a more manageable size. The most common, fundamental method for data synopsis is to use some kind of algorithm that reduces the data to a simpler set of descriptors. The mining algorithms can then either act solely on these descriptors, or reconstruct the data when necessary with minimal loss of the original values. Alternative methods sample subsets of the data, either chosen randomly or via some optimization algorithm in which the more valuable data are retained and less valuable discarded.

Arguably the simplest synopsis method is *sampling*, where a typically random selection of data are used for the analysis. Popular for stream mining is '*Reservoir sampling*' [4], which sees sampled data being drawn randomly from a reservoir of recent data, analogous to a time horizon. This method has been adapted to work over the moving windows necessary for stream data [8], and with modifications to apply a biased sampling [5] such that the sampling from the reservoir is smoother and the accuracy rate for queries can be maintained at acceptable levels. For example, a reservoir may have a 12-hour time horizon, but best results are obtained from the most recent hour, with an exponential fall off in utility of data records.

*Wavelets* are a popular method of decomposing data into a series of simpler basis functions that can then be used to recreate that data, as discussed by [4]. Different trends or features of the data are captured by different levels of the wavelet coefficients, and can be used to analyze specific parts of the data. The *Haar wavelet* is one of the easiest and simplest wavelet methods to use as it can easily be computed for any data series, and decomposed into a tree structure form – known as the *error tree* – from which the data are easily reconstructed. Due to the nature of how the tree is formed, if only part of the data needs to be reconstructed then only a subset of the tree needs to be utilized. Furthermore, only the coefficients that are sufficiently large need be retained as the smaller coefficients have a much lesser affect on the reconstructed result. The total number of coefficients for wavelet decomposition is equal to the length of the data record itself, so only the largest coefficients should be retained in order to make computational gains. However, it is not always the case that the coefficients with the largest absolute value are the best to keep, and metrics such as the mean and maximum square error should be employed to make an appropriate choice.

Another data summary method is that of *sketches* [25], where random projections are made to summarize the data with increasing complexity until the data can be reconstructed to a desired accuracy. In essence, this is a similar process to wavelets, and is typical of such summary methods that enable data to be reduced to a series of coefficients, a subset of which will reasonably approximate and reproduce the original data series to within some bounds. Further examples are *principal component analysis* (PCA) and *linear regression*, both of which are discussed by [23] in detail, and *discrete Fourier transforms* (DFT).

While these summary methods are well established, it is only more recently that they have been adapted for the stream data environment. Traditionally, such methods have been employed on static data sets and thus provided a well-defined solution. With stream data, the unbounded nature of the problem and the possibility of changing trends within the data require that coefficients be tracked and updated on a continuous basis. While not an enormous challenge, one difficulty encountered with these methods is that the coefficients being tracked may decrease in importance over time whereas other coefficients that have been approximated to zero and essentially "forgotten" may become important.

The *Streaming Pattern dIscoverR in multIple Time-series (SPIRIT)* algorithm [23] is a framework that discovers and summarizes correlations in data streams, and relies on auto-regression methods to determine principal components of observed variables and the weights of these variables. SPIRIT addresses the issue of changing weights of variables by periodically reassessing all components either when the system load is low, or the results begin to fall below some pre-determined error threshold. Algorithms such as this can be adapted to work for time-horizons of any duration, and so-called "forgetting factors" can be introduced to bias coefficients to favor more recent data within a window.

The greatest utility of the SPIRIT algorithm, however, is in its ability to perform change detection within streams. The algorithm is designed to monitor incoming data, model the trends within the data in as few coefficients as possible, and then raise an alert or flag when the nature of the data changes. Typically this would be when a new descriptor variable is found and needs to be tracked, or a new underlying trend is discovered in the data. This makes it a very powerful tool for event detection, which is fundamental to solar physics.

One of the most ubiquitous data summary methods is that of histograms, which summarize data by placing it into discrete "bins" of values within certain ranges, with a count of the population of each bin maintained at every step. Histograms are useful for summary statistical analyses of data, though have limitations as the statistics of data within the bins is lost, and interpolations across histogram bins can then introduce large errors. In stream data, wavelet-based methods [2, 23] can be used to dynamically maintain histograms, with the maintenance being performed on the frequency, rather than cumulative, distributions. Again this relies on dynamically maintaining the wavelet coefficients but, as previously mentioned, only the largest and most relevant coefficients need to be tracked.

Reducing and summarizing our large volumes of solar data will likely prove a critical component to analyzing it in the streaming sense. A significant portion of each data file is a series of unchanging or irrelevant pixels (particularly those not on the solar disk itself), and retaining those values is often a waste of resources. Reducing the data to summary values can eliminate unwanted data values from consideration and lighten the burden on the processing system. Sampling of the data can also reduce the computational load, particularly when we consider the typically high correlation between consecutive solar images. However, there are significant limitations to summary methods with solar imaging data. If we are interested in detecting, tracking and monitoring specific physical features then many of these summary methods will have limited use as the spatial and locality information of the image is lost. The data becomes statistically summarized and easily analyzed, but no information regarding which pixels are neighboring other pixels will be retained. These methods could best be employed for the statistical analyses of the data.

The ability of summary methods to be employed for change detection is also worth noting. In many cases the data is monitored to look for sudden events or changes that are out-of-the-ordinary. Given the unprecedented spatial and temporal resolution that we now face in the solar physics community, there may be small-scale events, frequencies or underlying patterns that until now remain undetected, simply because no one has sampled the data at the highest rate in order to look for them. A change detection algorithm would be ideal in this situation, monitoring the data over long periods of time, providing a baseline of "behavior" at the very highest available resolution, and then creating an alert to sudden, unusual, or unexpected trend changes within the data.

*D. Change Detection and Concept Drift*

Synopsis and summary methods excel at reducing large volumes of data, but they also lend themselves extremely well to change detection – that is, determining when some fundamental change has occurred in the data. This is an important tool for detecting new or anomalous events within data streams, or discovering sporadic underlying trends.

As discussed, the SPIRIT algorithm [23] is one such method designed to perform change detection and locate so-called "hidden variables" – that is, signals, trends or variables within a data stream not readily apparent but that account for sporadic changes in the nature of the streams. SPIRIT does this by creating a model of the behavior of a stream in as few variables as possible, regularly updating the model parameters to ensure it is within acceptable bounds. The framework tweaks the model parameters as needed, and adds or removes terms if the system becomes under or over-determined.

In addition to being able to detect sudden changes, many stream mining algorithms must also be robust enough to handle so-called *concept drift* in the data. As stream data arrives, it may evolve over some timescale, requiring new models, classifications, or clusters, in order to accurately represent the data. For clustering of data, the retention of micro-clusters [1] for example, is a good way to capture the evolution of a stream by taking well-placed snapshots in time and storing those for later offline analysis and merging into macro-clusters. The corresponding pyramidal time-frame regime is critical in guiding when the snapshots are recorded.

To specifically address changes and evolution in data streams, a "*velocity density method*" [3] has been devised that enables the construction of a density-based velocity profile of the data, analogous to kernel density estimation methods used in static data sets. The density of a region of data is the smoothed sum of the data created by a user-defined kernel, and a temporal window within some pre-determined time horizon is used to calculate the velocity density. The method creates two estimates, called the *forward* and *reverse time slice density estimate*s that measure the density functions for all spatial locations in the past and future windows, respectively. It's important to note that both functions use the same data points, but each assumes time flows in a different direction. Then, by looking at the rate of change of the density estimates - or *velocity* density - over the time horizon, changes in the stream can be both quantified and visualized. Integrating the velocity density over the entire spatial range of the data can lead to what is known as the *total evolution* of the data – a metric of the change of the nature of the data.

Again, the highly adaptable micro-clustering technique can be incorporated here to determine regions of "expansion", "contraction" or even stability in the data. When combined with appropriately chosen time horizons and decay rates in the windows, this method can be used to detect either short or long term changes in the data, though it is noted that it can be hard to tune the parameters to detect both. Thus prior to employing these methods, there needs to be a clearly defined set of goals in terms of the analysis performed, and the algorithms tuned accordingly.

Change detection and concept drift would seem to be important concepts to keep in mind in the context of stream mining of solar data. The Sun is extraordinarily dynamic on every timescale, from seconds to weeks. Thus if, say, the goal of stream mining the data is to find new classes of events that as yet remained undetected, then it will first be necessary to create models that capture and account for all of the current ongoing changes that we know of so as to detect new one. A framework like SPIRIT could likely excel in this respect, as the results given in [23] demonstrate the algorithm's ability to accurately model even fairly complex data streams with minimal variables, and then adapt and capture new trends as they appear. It is important to understand, however, the SPIRIT is designed for handling multiple streams from multiple sources, and in particular sensor data that returns a single discrete value (e.g. temperature, water flow) at each time increment. There would need to be an additional step of pre-processing the solar data to get that data into a more suitable form. The method would seem to be limited for morphological feature tracking, but should adapt well in searching for periodicities and statistical signatures.

*E. The Role of Databases*

Since the launch of the SOHO satellite in the 1990's, the solar physics community has come to rely on databases for storing and querying data. Each image returned by the spacecraft comes with it a set of metadata that provides information such as time, date, exposure time, wavelength, and image size, for example, as well as spacecraft information such as location, attitude, roll, temperatures etc. By querying this data, scientists are more easily able to get to the information they require.

In more recent years, supplemental databases have been formed from the SOHO data to catalog events such as coronal mass ejections[6] and flares. These data products are typically created offline during a follow-up pass over the data, often with manual analysis by a scientist. The more comprehensive Virtual Solar Observatory stores a wealth of information and metadata regarding solar features and events, with data drawn from a variety of sources. Clearly the community embraces the utility of databases, it is worth briefly reiterating that in addition to effectively mining future volumes of solar data, we must be sure to appropriately store all metadata produced.

In the light of the methods discussed in this paper, appropriate databases are essential for retention and later analysis of summarized data any results. Any adaptation of the algorithms mentioned here to mine solar data would require an appropriate database to store the results and enable offline queries. Depending on the types of features analyzed, database structures may need to store spatial information (latitude and longitude, for example) if physical morphological features are being mined, or a more traditional database structure for statistical or numerical parameters or coefficients. A full discussion of these database requirements is beyond the scope of this paper. This point serves only to emphasize that performing effective mining of the solar data streams is only half of the battle.

## VI. CONCLUSIONS

The NASA SDO mission presents a watershed moment in solar physics. The current and future volumes of data from this mission alone present both unprecedented challenge and opportunity for solar physicist to gain valuable new insight into the processes occurring on the solar surface and within the solar corona. However, to fully exploit these enormous increases in spatial and temporal resolution, and the incoming volume of data, the community must look beyond simple methods of analysis and data mining. Instead, they must consider the approach of other scientific and computational communities better versed in dealing with high-volume data streams. In this survey paper we have looked at several applicable categories of the methods and algorithms employed by computer scientists when mining high-volume or high-velocity data streams.

One of the most important analyses performed in solar physics is simply cataloging and recording features on the solar surface. With the new influx of ultra high-resolution data, physicists must now consider the possibility of cataloging small and shorter duration features that have previously remained under-sampled. For this, lessons can be taken from efforts in classification and clustering of data streams, where both known and unknown are recorded and grouped together according to the nature of the feature. Related to this is frequent pattern mining, where the most frequently recurring features are to be recorded. Central to these efforts, the proposed micro-clustering approach seems to hold great promise for its applicability to the solar data, though modification would certainly be needed, and care would need to be taken to ensure that the changing nature of features is fully accounted for.

Due to the unbounded nature of streams, the data cannot be analyzed in whole. Thus methods must adapt to either handle one data element at a time, or operate within some reasonable time horizon. For immediate event or feature detection, the latter would be preferable to take advantage of the advantage of high spatial and temporal resolution. However, long-term trends and recurring features may still play a crucial role, and thus consideration must be given to retaining memory of older features.

Synopsis methods are commonly used to allow long-term trends to be recorded and analyzed. Utilizing a synopsis method such as wavelets or PCA, the data can be represented by a vector of coefficients that are much smaller and thus computationally simpler to work with. An added bonus of synopsis methods is that these coefficients can be monitored for either short, long-term, or abruptly occurring changes.

Finally, any data, results, or synopses that are mined from data should be appropriately recorded and cataloged into a database. This database will likely incorporate elements of both traditional and spatial databases, as numerical and spatial data are recorded. While this discussion was beyond the scope of this study, it will prove to be a vital consideration for any stream mining algorithms applied to the data.

There exists no single one-stop solution for stream mining solar data. The nature of the images recorded, when combined with the spatial and temporal resolution, and the kinds of features that would need to be detected, have no direct analog outside of the field. Thus any successful mining efforts for this dataset will require a combination of several of the algorithms and tools mentioned here, and no doubt some novel approaches in order to combine them all for use in such a unique situation. The algorithms covered in this survey are simply those that offer the most obvious applicability to the problem, and are stated primarily to offer examples of the fundamental method that they represent.

There are numerous extensions to the problem that have not been considered in this survey. This study has placed a strong emphasis on SDO, given that the mission is currently alone in terms of its volume of data. However, complementary new mission such as IRIS, and even older missions like HINODE and STEREO, continue to return significant data volumes. Other active near-Earth missions such as GOES and ACE are concurrently monitoring solar and space weather activity near Earth, returning data in near real-time or at least on an ongoing basis. Finally, existing and planned ground-based observatories and surveys continue to add to our ever-growing the pool of solar data. An ideal, comprehensive model of the Sun's behavior would incorporate data from all of these sources. This would require an additional set of stream analysis methods not discussed here, such as stream joins [31] and load balancing [29, 6], as we look to combine multiple overlapping data records, arriving sporadically and threatening to overwhelm the system. Related to this, we have not discussed that SDO, for example, consists of three different instruments recording images in several wavelengths (or filters). Successfully combining this data would present further stream joining, classification and clustering challenges, to name a few.

---

[6] http://cdaw.gsfc.nasa.gov/CME_list/

One of the goals of this survey was to introduce the solar community to a paradigm outside of the normal scope of its study – namely stream processing. We have provided a very broad overview of the basic methods used for mining data streams, given examples of specific established algorithms, and stated the application and limitations of these approaches. However, this study has barely scratched the surface of a much larger problem that faces the whole astronomical community in coming years. Planned planetary, heliophysics and astrophysics missions are now able to return extraordinary data rates with modern technology. For example, the NASA Lunar Atmosphere and Dust Environment Explorer (LADEE) mission utilizes a revolutionary laser-based communication system that has been demonstrated to return up to *600 Megabits per second*. [10] Missions such as this will challenge scientists to reconsider their traditional methods of analyses and look to those employed by communities more used to dealing with such volume, velocity and variety of data.

ACKNOWLEDGMENT

The author wishes to thank Prof. Huzefa Rangwala (George Mason University) for guidance and discussion in compiling this survey. This work was supported by NASA.